\documentclass[aps,pra,twocolumn,groupedaddress,floats,showpacs]{revtex4}

\usepackage{amsmath}
\usepackage[dvips]{graphicx}
\usepackage[dvipdfm]{hyperref}

\newcommand{\s}{\sigma}

\newcommand{\ns}{\mathcal{N}_{\rm s}}

\newcommand{\ps}{\hat{\psi}^{\vphantom{\dagger}}}

\newcommand{\psd}{\hat{\psi}^{\dagger}}

\newcommand{\ket}[1]{|#1 \rangle }

\renewcommand{\vec}[1]{\mathbf{#1}}
\renewcommand{\vr}{\vec{r}}

\begin{document}

\title{Role of interactions in time-of-flight expansion of atomic clouds from optical lattices}
\author{Joern~N.~Kupferschmidt$^{1,2}$ and Erich J. Mueller$^1$}
\affiliation{$^1$Laboratory of Atomic and Solid State Physics,
Cornell University, Ithaca, NY 14853-2501, USA}
\affiliation{$^2$Dahlem Center for Complex Quantum Systems and Institut f\"ur Theoretische Physik, \\ Freie Universit\"at Berlin, 14195 Berlin, Germany}

\begin{abstract}
We calculate the effect of interactions on the expansion of ultracold atoms from a single site of an optical lattice.  We use these results to predict how interactions influence the interference pattern observed in a time of flight experiment.  We find that for typical interaction strengths their influence is negligable, 
yet that they reduce visibility near a scattering resonance.
\end{abstract}


\maketitle

\section{Introduction} 
One of the most important probes of cold atom systems is time-of-flight imaging. 
Turning off all trapping potentials, a cloud of cold atoms expands for tens of milliseconds, and an absorption image is taken. 
 In the far field limit, the resulting image can be directly interpreted as the momentum distribution of the original cloud, {\em if} interactions among the atoms can be neglected during the expansion.  Here we critically evaluate the validity of neglecting such interactions during the expansion from an optical lattice.

The question of how to interpret time-of-flight images is crucial.  These images have been used, for example, to distinguish the Mott insulating and superfluid phases \cite{greiner2002,spielman2008}.   
They have also been proposed as a tool to detect vortices in rotating condensates~\cite{goldbaum2009}, and are a crucial component of more sophisticated probes such as modulation spectroscopy \cite{modulation} and  Bragg/Raman spectroscopy \cite{bragg}.  

Interactions between cold neutral atoms are parameterized by the $s$-wave scattering length $a$, which is typically on the order of $5-15$nm \cite{hung2010,greiner2002}.  The scattering length is almost always very small compared to the distance $\lambda / 2 $ between sites in an optical lattice, $  \lambda/2\approx 426$nm, where $\lambda $ is the wavelength of the laser used to create the optical lattice~\cite{greiner2002}.  The scattering length can, however, begin to approach the size $\sigma_{\rm r}$ of the atomic states in one well.  For example  $\sigma_{\rm r}\approx 75$nm for a modest optical lattice with depth $V_0\sim 10 E_R$, where  $E_{\rm R} = \hbar^2(2\pi)^2/(2m\lambda^2)$  is the recoil energy of the lattice.  Thus when a few particles occupy a single site, their interactions are significant~\cite{hazzard2010}. Experiments have measured the resulting energy shifts~\cite{campbell2006},
and recently used them to study atom number statistics~\cite{will2010}. While these on-site interactions are important, by the time the wave-packets have expanded enough to overlap with neighboring sites, interactions are greatly attenuated.

Hence in our analysis we include interactions between atoms expanding from the same site, but neglect all inter-site interactions. Thus we are able to investigate whether interactions during the initial expansion period affect the interference image. Our estimate of the role of interactions is a lower bound; there may be further interaction effects during later stages of the expansion. Most importantly, interference effects could lead to strongly interacting high density regions at intermediate times~\cite{spielmanprivcomm}.

Within our approximation, the density profile of the many-body system during time of flight depends only on the $t=0$ wavefunction, and the time dependence of a cluster of particles expanding from a single site. In the next Sec.~\ref{single} we consider the expansion from a single site. Time-of-flight interferometry is considered in Sec.~\ref{interfere}. As a numerical example we consider the expansion of a two-dimensional, harmonically trapped cloud forming a superfluid in Sec.~\ref{example}. We summarize our results in Sec.~\ref{summary}.

In 2008, Gerbier et al. \cite{gerbier2008} reported the results of a very similar calculation, however they gave very few details.  More recently, Fang, Lee, and Wang \cite{fang} reported a complementary investigation, where they used a truncated Wigner approximation to investigate the role of interactions during time-of-flight expansion.  Restricting the expansion to one dimension (1D), they considered the dynamics of 10 atoms released from a 10-site optical lattice.  As we discuss in section~\ref{phase}, interactions play a much larger role in 1D expansion than in 3D, and Fang et al. consequently found nearly a factor of two attenuation of the central Bragg peak compared to the noninteracting gas.  Using very similar parameters, we find that interactions during 3D expansion only lead to a 5\% reduction in the amplitude of the central Bragg peak.

\section{Single Site Expansion}\label{single}
\subsection{Statics}
An optical lattice is typically modeled as a potential of
  the form
\begin{equation}
V(x, y, z) \,= \, V_0  [ \sin^2(k x) + \sin^2(k y) + \sin^2(k z)]
\end{equation}
where $ k = 2 \pi / \lambda$.
Near the local minima one may
approximate the sinusoidal as a harmonic potential
$ V_{\rm eff} = m \omega_{\rm r}^2 \vr^2 / 2 $,
with small oscillation frequency
$\omega_{\rm r}  = \,  (\hbar k^2 / m)  \sqrt{ V_0 / E_{\rm R}}$.
 The single-particle ground state in this potential is a Gaussian
%
\begin{equation}\label{spwf}
   \phi_{1,i}(\vec r)
   \, = \, \frac{1}{(\pi \sigma_{\rm r}^2)^{3/4}}
   \exp \left[ - \frac{ ( \vec{r} - \vec{r}_i)^2}{2 \sigma_{\rm r}^2}
\right] ,
\end{equation}
where $\sigma_{\rm r}^2  =  \hbar/ m \omega_{\rm r}$.
We model the interaction among the particles as a contact interaction,
\begin{equation}
   \hat H_{\rm int}
   \, = \,
  \frac{ g}{2} \int d^3 r  \psd(\vr) \psd(\vr) \ps(\vr) \ps(\vr),
\end{equation}
where $g=   4 \pi \hbar^2 a/m$.

We are concerned about how these interactions modify the few-body wavefunction on a single site, and how this influences the time-of-flight expansion.  As long as $a\ll\sigma_{\rm r}$, the effects of interactions are captured by a Gaussian variational ansatz
\begin{align}
\label{eq:phi} 
  \Phi_{N}(\{\vr_{\alpha}\}) 
  =&  e^{i \xi_N} \phi_{\rm cm}( \vec{r}_{\rm cm} ) 
      \prod_{\alpha < \beta}\phi_{N}( \vr_{\alpha} - \vr_{\beta} )
 \\ 
  \phi_{\rm cm}(\vr) 
   =&  \frac{ 1}{ ( \pi \sigma_0^2 )^{d/4}} 
   \exp\left[ - \left( \frac{ N }{2 \sigma_0^2} -  \frac{ i \beta_0}{ N }\right)  \vr^2 \right]  \nonumber  \\ 
   \phi_{N} (\vr)  
   =& \frac{ 1}{ (\pi \sigma_N^2)^{ d/4} } 
   \exp\left[- \left( \frac{ 1}{2 N \sigma_N^2 } -  i N \beta_N \right) \vr^2 \right] \nonumber
\end{align} 
where we have introduced the center of mass $ \vec{r}_{\rm cm} =(1/N) \sum_{\alpha = 1}^{N} \vec{r}_i $ and generalized to arbitrary spatial dimension $d$. 
For $N = 1 $, Eq.\ (\ref{eq:phi}) reduces to the harmonic oscillator 
wavefunction $ \phi_{1}(\vec r)$. 
We will use this wavefunction as a time-dependent variational ansatz to describe the dynamics, hence we have introduced the parameters $\beta_0 $ and $\beta_N $ which are nonzero only if the cluster is expanding or shrinking.

For convenience we will work with dimensionless quantities, using units where 
$\hbar  = m= \lambda / 2 = 1$. 

\subsection{Dynamics} 
One produces a variational estimate of the dynamics by minimizing the action 
\begin{align}
\label{eq:action}
 S & = \int dt \int \prod_{\alpha} d^d r_{\alpha} 
   \left\{ \frac{i}{2} \left[ \phi^*  \frac{ \partial \phi} { \partial t}  
  - \left( \frac{ \partial  \phi^* }{ \partial t} \right) \phi \right] 
   \right. \\   \nonumber& \left. 
  -  \phi^* \left[ 
  - \frac{1}{2} \sum_{\alpha} \left( \frac{ \partial^2}{ \partial \vr_{\alpha}^2 } 
  + \frac{\vr_{\alpha}^2}{\s_r^2 } \right)
  + \sum_{\alpha<\beta} g \delta^{(3)}(\vr_{\alpha} - \vr_{\beta}) \right] \phi \right\} . 
\end{align}
We use the trial wave-function in Eq.~(\ref{eq:phi}), for which the spatial integrations can be performed analytically, and allow all variational parameters to be arbitrary functions of time.
A similar approach has been used to describe the role of interactions in the dynamics of a harmonically trapped BEC, where the atom number is much  larger~\cite{perez-garcia1996}. Minimizing the action leads to a second order differential equation for the width $\sigma_N$,
\begin{eqnarray} 
\label{eq:s_evolution}
   \s_N^3 \frac{\partial^2 \s_N}{ \partial t^2} 
   & = & 
   1  - \frac{\s_N^4}{ \s_r^4} +   \frac{ N g }{(  2 \pi)^{d/2} }  \s_N^{2 - d},\\\nonumber
  \beta_N &= & \frac{1}{2 \s_N}  \frac{\partial  \s_N }{\partial t}.
\end{eqnarray}
The center-of-mass width $\sigma_0$ obeys the same equation, but with $g=0$.  Note that in the noninteracting limit all
$N$ dependence drops out.  The sole contribution from the optical lattice is the term $\s_N^4/\s_r^4$.  During time-of-flight expansion, the optical lattice as well as the  harmonic trapping potential are removed, and this term no longer appears in the equations of motion.

At time $t=0$ we set $\beta=0$, and take $\sigma$ to be given by the static solution with $\partial_t \sigma = 0$.  Analytic solutions to the resulting algebraic equation can only be found when $d=2$, where
\begin{equation}
  \s_N(0, d = 2)  = \sqrt{ 1 + \frac{N g }{ 2 \pi} } \s_r .
\end{equation}
The center-of-mass width is simply
$\s_0^2 =  \s_r^2$ in all dimensions.

In two dimensions we can analytically integrate the equations of motion,
\begin{eqnarray}
   \s_0(t)  & =& \sqrt{ \s_r^2 +  t^2 / \s_r^2 }  \\
   \s_N(t, d=2 ) & = & \left( 1 + \frac{ N g }{ 2 \pi} \right)^{1/4} \s_0(t),   \\
  \beta_0(t)=\beta_N(t,d=2) 
   & =&
  \frac{ 1 }{ 2 } \frac{ t}{ \s_r^4 + t^2 }.
\end{eqnarray}
The expressions for $\s_0$ and $\beta_0$ apply in all dimensions.

\subsection{Phase accumulation}\label{phase}

The phase of the expanding cluster is crucial for determining the observed interference pattern.  In terms of $\sigma_N$, one finds
\begin{align} 
\label{eq:xi_evolution}
  \frac{\partial \xi_N}{ \partial t}
   = & 
  -  \frac{ N d}{ 2 \s_0^2 } 
 \\   \nonumber&
  -  \frac{ N -1 }{2} \left[\frac{ d} { \s_N^2 } - \frac{d}{ \s_0^2 } +  ( d + 2 ) \frac{ N g }{ 2}   \left( \frac{ 1 }{ 2 \pi \s_N^2 }\right)^{d/2} \right].
\end{align}
The contribution in square brackets arises from the interactions. 
Interactions increase the width of the 
initial state, which reduces the contribution of the kinetic energy and retards the phase relative to the noninteracting expansion.  This should be contrasted with the contribution from the potential energy as well as the direct interparticle interactions themselves, which increase the energy and advance the phase. To determine the net sign of the interaction correction is thus not straightforward. In particular,  $\s_N $ is generally 
larger than $ \s_0$ and the quantity in square brackets does not have a definite sign.

We produce a rough estimate of the phase accumulated by replacing $\sigma_N$ with $\sigma_0$ in this expression.  The interaction contribution to the phase will then scale as
\begin{equation}\label{xi1}
 \xi_{\rm int}\propto \int_0^t  d t \frac{1}{ \s_0(t)^d }  
   \, = \,  
  \left \{ \begin{array}{ll} 
  \s_r  \mbox{arcsinh} \left( \frac{t}{ \s_{\rm r}^2} \right)  &  d = 1 
  \\ 
   \mbox{arctan}\left( \frac{t}{\s_{\rm r}^2} \right) &  d = 2 
  \\ 
  \frac{ t }{ \s_{\rm r}  \sqrt{ \s_{\rm r}^4 +  t^2 } } &  d= 3
  \end{array} \right. 
\end{equation}

Whereas the contribution is logarithmically divergent in the one-dimensional case, it very quickly reaches a finite value in the two-dimensional as well as the three-dimensional case. This indicates that the influence of interactions is confined to the very beginning of the time-of-flight expansion, 
$t \lesssim \sigma_{\rm r}^2$, essentially corresponding to the time required for the cluster to expand to less than twice its initial size, where $\sigma(t)^2 \approx 1/2$.  
(Recall, we are using units where lengths are measured in terms of the lattice spacing and times, up to numerical constants, in units of the inverse recoil energy.)  
Typically, this means that for $d=2$ or $d = 3$ interactions become irrelevant well before the clusters overlap.  Conversely, interactions between clusters can not be neglected during $d=1$ expansion.

The non-interacting contribution to the phase is
\begin{eqnarray} 
 \xi_0(t) & = & - \frac{N d}{2} \arctan{ \frac{ t }{ \s_0^2(t)}}.
\end{eqnarray}
In two dimensions, where we have analytic expressions for $\sigma_N(t)$, we further find
\begin{equation}
  \xi_N(t , d = 2 ) 
  =  
  - \left[ 1 + ( N - 1) \sqrt{ 1 + \frac{ N g}{ 2 \pi } } \right] 
  \arctan{\frac{ t }{ \s_0^2(t)}}.
\end{equation}
The fact that interactions only modify the prefactor is a reflection of the scaling symmetry of the expanding cloud in $d=2$.

\section{Time-of-flight images} 
\label{interfere}

Having calculated the expansion dynamics of a single cluster of particles, we now explore the consequences for the 
atom density seen in a time-of-flight expansion experiment.  Neglecting correlations between sites, we assume that the initial state can be written as a generalization of the standard Gutzwiller Ansatz,
\begin{equation}\label{ansatz}
  | \Psi  \rangle 
  \, = \, 
  \bigotimes_{i = 1}^{\ns} \left( 
  \sum_{n=0}^{\infty} f_{i,n} \int d^{3 n} r \,  
  \Phi_{n}(\{\vr_{\alpha} - \vec{R}_i\})  \, \ket{ \{ \vr_{\alpha}\}}
  \right) 
\end{equation}
where $i$ runs over all $\ns$ lattice sites $\vec{R}_i$, $n$ is the number of particles on a given site, and 
\begin{equation} 
  \ket{ \{\vr_{\alpha}\}}  
  \, = \, 
  \frac{ 1}{ \sqrt{ n!} } \psd(\vr_1) \psd(\vr_2) \ldots \psd(\vr_n) \ket{0}.
\end{equation}
The state is normalized when the norm of the $f$-vector is one,
\begin{equation}
  \sum_{\alpha = 0 }^{\infty} |f_{i,\alpha}|^2  = 1 .
\end{equation}
The $n$-particle wavefunction on site $i$,  $\Phi_{n}(\{\vr_{\alpha} - \vec{R}_i\})$, is given by Eq.~(\ref{eq:phi}).

Within our approximation, where we neglect interactions between atoms on different sites, the time evolution of Eq.~(\ref{ansatz}) simply amounts to separately time evolving each cluster, as described in Sec.\ \ref{single}.  The resulting density profile is
\begin{align} 
\label{eq:density}
   \langle \Psi | \hat \psi^{\dagger} ( \vr) \hat \psi(\vr)& | \Psi \rangle 
    =   
   \sum_{i=1}^{\ns} \sum_n  n | f_{i,n}|^2  | \phi_{n}^c ( \vr - \vec{R}_i )|^2 & 
 \\   
   + \sum_{i=1}^{\ns} \sum_{k \neq i }  
   & \left( \sum_n  \sqrt{n} f_{i,n}^* f_{i,n-1} \phi_{n,n-1}^{c*} (\vr-\vec{R}_i) \right)   
   \nonumber \\  
   \times & 
 \left(  \sum_m \sqrt{m} f_{k,m} f_{k,m-1}^* \phi_{m,m-1}^{c}  (\vr-\vec{R}_k) \right) \nonumber
\end{align}
with 
\begin{align}
  |\phi_{n}^c (\vr)|^2 
   =& 
   \int \Pi_{\alpha = 2}^{n} d^3 r_{\alpha} 
   | \Phi_{n}( \vr, \vr_2 ,\ldots, \vr_{n})|^2 
     \\ 
  \phi_{n,n-1}^c (\vr)  
   =&
   \int \Pi_{\alpha = 2}^{n} d^3 r_{\alpha}  
   \Phi_{n} ( \vr, \vr_2, \ldots ) \Phi_{n-1}^*( \vr_2 ,  \ldots).\nonumber
\end{align}
In the noninteracting limit both contractions reduce to the noninteracting single particle wavefunction, so that
\begin{eqnarray} \label{cond}
 |\phi_{n}^c (\vec{r})|^2 & = & | \phi_1(\vr)|^2 
\nonumber \\ 
\phi_{n,n-1}^c (\vec{r})  & = &  \phi_1(\vr).
\end{eqnarray}
In this case one can write a more readily interpretable expression for the density \cite{goldbaum2009}, 
\begin{align} 
  \label{eq:rho}
  \langle \Psi_{\rm ni} | \hat{n}(\vr) | \Psi_{\rm ni} \rangle 
  = &  |\phi_1(\vr)|^2  \left[ ( N - N_{\rm c} ) + | \Lambda ( \vr) |^2 \right]
   \\
  \label{eq:lambda}
  \Lambda ( \vr)  
  = & \sum_{i = 1}^{\ns}  \alpha_i e^{ - i \beta_0 ( \vr \cdot \vec{R}_i  - \vec{R}_i^2 ) }.
\end{align}
Here $N $ is the total number of particles in the lattice.  
$|\phi_1(\vr)|^2$ is a simple gaussian with 
width $ \sigma_0(t)/\sqrt{2}$. Corrections to the 
featureless gaussian peak, $N_{\rm c}$ and $\Lambda(\vr) $, signal the 
presence of superfluid order in the system. $N_{\rm c} $ is the condensed
 number of particles, whereas $\alpha_{i} $ is the expectation value of 
the annihilation operator 
on site $i$ and thus 
the superfluid order parameter in the system.
\begin{align}
 \label{eq:sforder}
  \alpha_i =& \langle \hat a_i \rangle = \sum_{n} \sqrt{n} f_{i,n}  f_{i,n-1}^* \\
  N_{\rm c} =&  \sum_{i = 1}^{ \ns} | \alpha_i |^2 \label{nc}
\end{align}
Gerbier {\em et al.}~\cite{gerbier2008} have pointed out 
that in Eq.\ (\ref{eq:lambda}) above, 
for experimentally relevant expansion times on the order of tens of 
milliseconds, it is necessary to keep the 
Fresnel like terms quadratic in $\vec{R}_i$. 
In the absence of the Fresnel terms, the shape of Bragg peaks 
is simply the Fourier transform of the superfluid order parameter.

Here  we go beyond the approximations in Eqs.~(\ref{cond})-(\ref{nc}), and include the effects of interactions on the expansion.  These interactions have two effects.  First they broaden each of the expanding clusters.  This broadens the incoherent background, but it also reduces the contrast of the Bragg peaks.  This latter effect occurs because of the reduced overlap between the expanding clusters with different numbers of particles.  Second, the interactions introduce a nonlinear phase difference between the different particle number clusters.  This dephasing further reduces the contrast of the Bragg peaks.

The broadening of the incoherent background is quantified by 
\begin{equation}
   |\phi_{N}^c (\vec{r})|^2 
   = 
   \frac{ 1}{ ( \pi \s_{N,\rm eff} )^{d/2} } \exp[ - \frac{ \vr^2 }{ \s_{N,\rm eff}^2} ], 
\end{equation}
where 
\begin{equation}
 \s_{N,\rm eff}^2 =  \frac{ (N -1 ) \s_N^2 + \s_0^2 }{ N } .
\end{equation}
Clearly $\s_{N,\rm eff} > \s_0$, reflecting the larger size of the interacting cluster. 

The influence of interactions on the Bragg peaks is quantified by identifying the difference from the noninteracting wavefunction $\phi_1(\vr)$, 
\begin{equation}
  \phi_{n,n-1}^c (\vec{r})  
   =   
  \phi_1(\vr) e^{ i \delta \xi_N} |\delta N_{N}| \exp[ \vr^2 \delta s_N]. 
\end{equation}
The overall phase $\delta\xi_N$, the width $(\sigma_0^{-2}-\delta s)^{-1/2}$, as well as the corresponding prefactor $|\delta N_{N} |$, affect the peaks.  The expressions for each of these terms are complicated, with 
\begin{align}
 \label{eq:deltaxi}
 \delta \xi_N& =   \xi_{N} -  \xi_{N-1} + \arg[\delta N_N]  - \xi_0
  \\ 
 \delta N_{N} 
  &= 
  \left( \frac{ 2 \s_{N} \s_{N- 1} }{ \s_{N}^2  + \s_{N-1}^2 - 2 i \s_N^2 \s_{N-1}^2 (\beta_{N} - \beta_{N-1} )} \right)^{(N-1)d/2}  
  \nonumber \\ & \times 
 \left[\frac{ 2 N \s_N \s_0 }{( 2 N  - 1 ) \s_N^2 +  \s_0^2  - 2 i \s_0^2 \s_N^2 ( \beta_{N} - \beta_0 ) }\right]^{d/2}
  \\ 
 \delta s_{N}& = \frac{N-1}{N \s_N^2}  \left[ \s_N^2  - \s_0^2 - 2 i  \s_0^2 \s_N^2 ( \beta_0 - \beta_N ) \right]  
\nonumber \\ 
 \times &
  \frac{ 2 N \s_N^2 + ( N - 2) \s_0^2( 1  - 2 i \s_N^2 \beta_N )}{ ( 2 N - 1)  \s_N^2  + (N-1) \s_0^2  -   2 i  \s_0^2 \s_N^2 [ (N-1) \beta_N - \beta_0 ]}.
\end{align}
In the noninteracting case the only phases contributing 
to the interference come from the terms $f_{i,n} f_{i,n-1}^* $ [cf.~Eq.~(\ref{eq:sforder})]. Here there are additional contributions as given by Eq.\ (\ref{eq:deltaxi}). 

\section{Numerical Example}
\label{example}

To illustrate our results, we consider a two dimensional harmonically trapped gas of $^{87}$Rb in optical lattices with $V_0=10.7, 8, 5 E_R$, yielding $\nu_r \approx  20.3, 17.5,  13.9$kHz. 
For each of these lattice depths we use a different scattering length, taking 
 $a = 5, 15.5, 50$nm, to give the same ratio $ U / 4 t \approx 15.5 $, well on the superfluid site of the superfluid-Mott transition~\cite{jaksch}. Adjusting the chemical potential $\mu_0 $ in the center of the trap to obtain the same total number of particles thus yields identical initial  states. 
To find the initial $f_{i,n}$'s of Eq.~(\ref{ansatz}) we solve the discrete variational Gutzwiller problem, minimizing $\langle \Psi_G| H_L |\Psi_G\rangle$, with
\begin{align}
H_L =& -t \sum_{\langle ij\rangle} (\hat{a}_i^\dagger \hat{a}_j + \hat{a}_j^\dagger \hat{a}_i) +\sum_i U \hat{n}_i(\hat{n}_i-1)-\mu \hat{n}_i  \\
|\Psi_G\rangle =&\prod_i \sum_n f_{i,n} \frac{(\hat{a}_i^\dagger)^n}{\sqrt{n!}}|{\rm vac}\rangle,
\end{align}
where $\langle i,j\rangle$ denotes nearest neighbor sites, $\hat{a}_i$ annihilates a boson at site $i$, $ \hat{n}_i = \hat{a}_i^{\dagger} \hat{a}_i $, and $t$ and $U$ are extracted from the non-interacting Wannier wavefunctions~\cite{jaksch}.  The corrections to $U$ from using the many-body wavefunctions on each site are very small 
at this lattice depth, and the corrections to $t$ are at most 10\% 
\cite{hazzard2010}.  We take the expansion to be three dimensional, treating the individual wells as spherically symmetric.

We produce initial conditions by solving Eq.~(\ref{eq:s_evolution}) with the conditions $\partial_t\sigma=0$ and $\beta=0$. Starting from these initial conditions, we numerically integrate Eqs.\~(\ref{eq:s_evolution}) and (\ref{eq:xi_evolution}). We then plot the densities, Eq.~(\ref{eq:density}).  

\begin{figure}
\includegraphics[width=80mm]{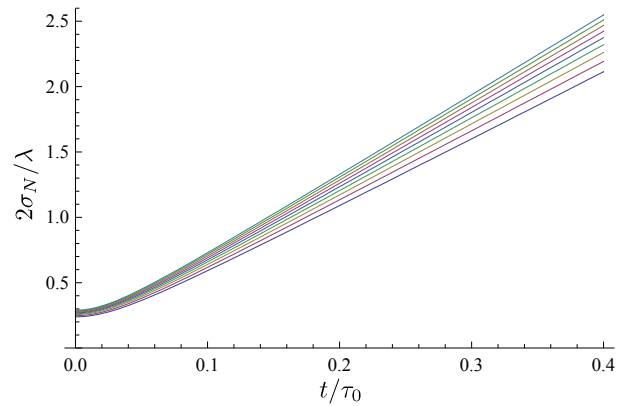}
\caption{\label{fig:widths} (Color online) Initial evolution of the widths of different number states $N = 1 $ through $10$ for $a = 50$nm and $V_0 = 5 E_{\rm R} $. The smallest widths are reached for $ N = 1$ (lowest curve), the largest for $ N = 10$ (highest curve). Interactions cause states with larger particle number to broaden and expand faster. Times are measured in units of $\tau_0 = m ( \lambda / 2 )^2 / \hbar$, which is approximately $0.25$ms for $^{87}$Rb. The range plotted is much shorter than a typical time-of-flight experiment. All subsequent expansion is ballistic.}
\end{figure}

Figure~\ref{fig:widths} shows the time evolution of the widths $\sigma_N$ of the clusters expanding from sites with different particle numbers.  As one can see, and as discussed in Sec.~\ref{single}, the expansion very quickly becomes ballistic.

\begin{figure}
\includegraphics[width=80mm]{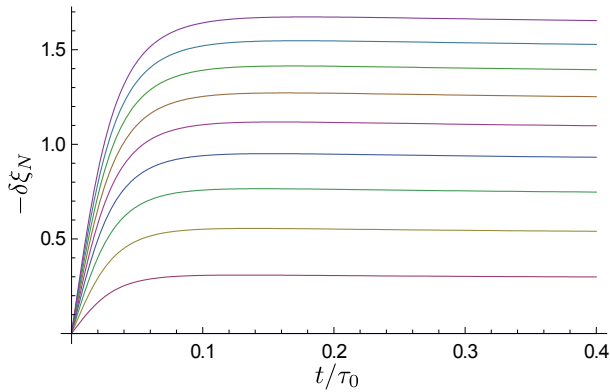}
\caption{\label{fig:phases} (Color online) Overall phase difference between wavefunctions with particle occupation differing by one [see Eq.(\ref{eq:deltaxi})], plotted for $ a = 50$nm and $V_0 = 5 E_{\rm R}$. Interactions do not affect the lowest (blue) curve, corresponding to $ n = \{0, 1\}$, but progressively affect the higher pairs $n = \{1, 2\},..., \{ 9,10\}$, yielding the largest effect for $n = \{9,10\}$, the top (violet) curve. The effect is approximately linear in $N$.
}
\end{figure}

Figure~\ref{fig:phases} shows the time evolution of the phase differences $- \delta \xi_N$, see Eq.~(\ref{eq:deltaxi}).  One sees that when $a=50$nm the phase difference between clusters of different particle numbers are on the order of $2\pi$, and hence the interference pattern will be influenced by the interactions. For typical $^{87}$Rb parameters, $ a = 5$nm, the phase difference is correspondingly smaller.

\begin{figure}
\includegraphics[width=85mm]{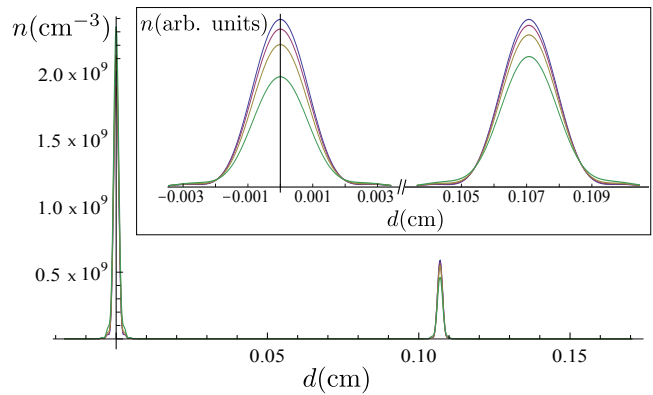}
\caption{\label{fig:densities} (Color online) Density of a nonrotating atomic cloud expanded for $100$ms, with distance $d$ measured from its center along a lattice direction. The interference peaks are clearly visible and remain so even in the strongest interacting limit considered here. Plotted are $ a = 0$nm (blue), $5$nm (violet), $15.5$nm (brown), and $50$nm (green). These progress from top to bottom in the inset and in the satellite peak at larger $d$, but bottom to top in the central peak. The inset shows the central and the first satellite peak with adjustment for different expansion velocities, as described in the main text. }
\end{figure}

Figure~\ref{fig:densities} shows cuts through illustrative density images along the lattice direction after a  100ms time-of-flight. This time was chosen to minimize distortions from Fresnel terms~\cite{gerbier2008}.  Weak interactions, $a=5$nm, have negligable effect on the image.  While stronger interactions $a=50$nm begin to reduce the amplitudes of the interference peaks, the peaks remain clearly visible. 

Comparing expansion images at fixed $U/ 4 t$ and fixed time of flight results in an interesting structure. In order to have the same $ U / 4 t $, the initial wavefunctions in the stronger interacting case must be larger, resulting in a slower initial expansion. Therefore in Fig.\ \ref{fig:densities} we see that the central peak is larger for stronger interactions, while the satellite peak is smaller. Using our intuition from the noninteracting expansion, one can think of this effect as being due to the envelope of the Bragg peaks which falls off on a scale inversely proportional to the size of the initial Wannier states. 
In the inset of Fig.\ \ref{fig:densities} we normalize out this effect by multiplying with the inverse envelope, $ ( \pi \sigma_0^2(t))^{3/2} \exp[ \vec{r}^2 / \sigma_0^2(t)]$. Taking this normalization into account, interactions reduce the amplitude of the interference peak. The reduction is about 5\% for $a = 5$nm, rising to 33\% for $a = 50$nm.  As shown in the inset of Fig. 3, the reduction is greater for the central Bragg peak than for the first satellite peak.

There are a number of ways of increasing the importance of the interactions during time of flight.  
For example, changing the geometry of the lattice sites influences how the cloud expands and how long interactions remain relevant:  the expansion from needle shaped sites is predominantly in the $x$-$y$ plane, and the 2D scaling in Eq.~(\ref{xi1}) approximately holds.  Additionally 
we have studied what happens when one suddenly increases the scattering length while releasing the atoms from the optical lattice.  This allows one to independently control the size of the initial Wannier states and the scattering length.  In the expansion shown in Fig.~\ref{fig:densities}, where $t/U$ was fixed while changing $a$, the initial Wannier states were larger when $a$ was made larger.

Starting from the equilibrium state with $a=5$nm and $V_0=10.7 E_R$, we investigate the expansion for $a=15.5$nm and $a=50$nm.  Although we do not show the results here, we find that the suppression of the central interference peak is roughly a factor of $1.5$ greater than what is seen in the inset of Fig.~\ref{fig:densities}.  

\section{Summary and Discussion}
\label{summary}

 We have considered the effect of two-particle interactions on the time of flight images of 
cold atoms on optical lattices.  We show that on-site interactions can be important for these images, but argue that one can neglect the interactions between atoms on different sites.

We find that even if interactions are increased by a factor of ten from their normal strength, no qualitatively new features appear in the time of flight images.  However, the quantitative size of the peaks is sensitive to the interactions. Given the wide tunability achievable by employing Feshbach resonances~ \cite{pollack}, it is conceivable that experiments can study the role of interactions during time of flight. The analysis presented here will fail when the scattering length $a $ becomes comparable to the size of the Wannier state $\sigma_r$. 


Conceptually it is worth noting that in many electronic mesoscopic systems a situation markedly different from the one here is encountered. There the dynamics is determined by impurity scattering or scattering off system boundaries, and at low temperatures are not affected by the interaction. Consequently there exists a regime in which interactions effectively only add additional phases to the relevant propagation amplitudes. Such a regime is not identifiable in the system we considered here. Instead, we find that whenever the interactions produce relevant phases, they also perturb the dynamics. 

\section{Acknowledgements}
We thank Daniel Goldbaum,  Kaden Hazzard and Ian Spielman for discussions. This work was supported by NSF under Grant No. PHY-0758104, through the Cornell Center for Materials Research, and the Alexander von Humboldt Gesellschaft.


\begin{thebibliography}{7}
\expandafter\ifx\csname natexlab\endcsname\relax\def\natexlab#1{#1}\fi
\expandafter\ifx\csname bibnamefont\endcsname\relax
  \def\bibnamefont#1{#1}\fi
\expandafter\ifx\csname bibfnamefont\endcsname\relax
  \def\bibfnamefont#1{#1}\fi
\expandafter\ifx\csname citenamefont\endcsname\relax
  \def\citenamefont#1{#1}\fi
\expandafter\ifx\csname url\endcsname\relax
  \def\url#1{\texttt{#1}}\fi
\expandafter\ifx\csname urlprefix\endcsname\relax\def\urlprefix{URL }\fi
\providecommand{\bibinfo}[2]{#2}
\providecommand{\eprint}[2][]{\url{#2}}

\bibitem[{\citenamefont{Greiner et~al.}(2002)\citenamefont{Greiner, Mandel,
  Esslinger, H{\"a}nsch, and Bloch}}]{greiner2002}
\bibinfo{author}{\bibfnamefont{M.}~\bibnamefont{Greiner}},
  \bibinfo{author}{\bibfnamefont{O.}~\bibnamefont{Mandel}},
  \bibinfo{author}{\bibfnamefont{T.}~\bibnamefont{Esslinger}},
  \bibinfo{author}{\bibfnamefont{T.}~\bibnamefont{H{\"a}nsch}},
  \bibnamefont{and} \bibinfo{author}{\bibfnamefont{I.}~\bibnamefont{Bloch}},
  \bibinfo{journal}{Nature} \textbf{\bibinfo{volume}{415}}, \bibinfo{pages}{39}
  (\bibinfo{year}{2002}).

\bibitem[{\citenamefont{Spielman et~al.}(2008)\citenamefont{Spielman, Phillips,
  and Porto}}]{spielman2008}
\bibinfo{author}{\bibfnamefont{I.~B.} \bibnamefont{Spielman}},
  \bibinfo{author}{\bibfnamefont{W.~D.} \bibnamefont{Phillips}},
  \bibnamefont{and} \bibinfo{author}{\bibfnamefont{J.~V.} \bibnamefont{Porto}},
  \bibinfo{journal}{Phys. Rev. Lett.} \textbf{\bibinfo{volume}{100}},
  \bibinfo{pages}{120402} (\bibinfo{year}{2008}).

\bibitem[{\citenamefont{Goldbaum and Mueller}(2009)}]{goldbaum2009}
\bibinfo{author}{\bibfnamefont{D.~S.} \bibnamefont{Goldbaum}} \bibnamefont{and}
  \bibinfo{author}{\bibfnamefont{E.~J.} \bibnamefont{Mueller}},
  \bibinfo{journal}{Phys. Rev. A} \textbf{\bibinfo{volume}{79}},
  \bibinfo{pages}{021602(R)} (\bibinfo{year}{2009}).


\bibitem{modulation}
T.~St\"oferle, H.~Moritz, C.~Schori, M.~K\"ohl, and T.~Esslinger, Phys. Rev. Lett. {\bf 92}, 130403 (2004).

\bibitem{bragg}
D.~Cl\'ement, N.~Fabbri, L.~Fallani, C.~Fort, and M.~Inguscio, Phys. Rev. Lett. {\bf 102}, 155301 (2009).


\bibitem[{\citenamefont{Hung et~al.}(2010)\citenamefont{Hung, Zhang, Gemelke,
  and Chin}}]{hung2010}
\bibinfo{author}{\bibfnamefont{C.-L.} \bibnamefont{Hung}},
  \bibinfo{author}{\bibfnamefont{X.}~\bibnamefont{Zhang}},
  \bibinfo{author}{\bibfnamefont{N.}~\bibnamefont{Gemelke}}, \bibnamefont{and}
  \bibinfo{author}{\bibfnamefont{C.}~\bibnamefont{Chin}},
  \bibinfo{journal}{Phys. Rev. Lett.} \textbf{\bibinfo{volume}{104}},
  \bibinfo{pages}{160403} (\bibinfo{year}{2010}).

\bibitem[{\citenamefont{Hazzard and Mueller}(2010)}]{hazzard2010}
\bibinfo{author}{\bibfnamefont{K.~R.~A.} \bibnamefont{Hazzard}}
  \bibnamefont{and} \bibinfo{author}{\bibfnamefont{E.~J.}
  \bibnamefont{Mueller}}, \bibinfo{journal}{Phys. Rev. A}
  \textbf{\bibinfo{volume}{81}}, \bibinfo{pages}{031602(R)}
  (\bibinfo{year}{2010}).

\bibitem{campbell2006}
  G.~K.~Campbell, J.~Mun, M.~Boyd, P.~Medley, A.~E.~Leanhardt, L.~G.~Marcassa, D.~E.~Pritchard, W.~Ketterle, Science, {\bf 313}, 649 (2006).

\bibitem{will2010}
  S.~Will, T.~Best, U.~Schneider, L.~Hackerm\"uller, D.\nobreakdash-S.~L\"uhmann, and I.~Bloch, 
  Nature, {\bf 465}, 197 (2010).

\bibitem{spielmanprivcomm}
 I.~B.~Spielman, private communication.
 
 

\bibitem[{\citenamefont{Gerbier et~al.}(2008)\citenamefont{Gerbier, Trotzky,
  Foelling, Schnorrberger, Thompson, Widera, Bloch, Pollet, Troyer,
  Capogrosso-Sansone et~al.}}]{gerbier2008}
\bibinfo{author}{\bibfnamefont{F.}~\bibnamefont{Gerbier}},
  \bibinfo{author}{\bibfnamefont{S.}~\bibnamefont{Trotzky}},
  \bibinfo{author}{\bibfnamefont{S.}~\bibnamefont{Foelling}},
  \bibinfo{author}{\bibfnamefont{U.}~\bibnamefont{Schnorrberger}},
  \bibinfo{author}{\bibfnamefont{J.~D.} \bibnamefont{Thompson}},
  \bibinfo{author}{\bibfnamefont{A.}~\bibnamefont{Widera}},
  \bibinfo{author}{\bibfnamefont{I.}~\bibnamefont{Bloch}},
  \bibinfo{author}{\bibfnamefont{L.}~\bibnamefont{Pollet}},
  \bibinfo{author}{\bibfnamefont{M.}~\bibnamefont{Troyer}},
  \bibinfo{author}{\bibfnamefont{B.}~\bibnamefont{Capogrosso-Sansone}},
  \bibnamefont{et~al.}, \bibinfo{journal}{Phys. Rev. Lett.}
  \textbf{\bibinfo{volume}{101}}, \bibinfo{pages}{155303}
  (\bibinfo{year}{2008}).
 
 \bibitem{fang}
 S. Fang, R.-K. Lee, and D.-W. Wang, arXiv:0910.1518 (2009).

\bibitem[{\citenamefont{Perez-Garcia et~al.}(1996)\citenamefont{Perez-Garcia,
  Michinel, Cirac, Lewenstein, and Zoller}}]{perez-garcia1996}
\bibinfo{author}{\bibfnamefont{V.~M.}~\bibnamefont{Perez-Garcia}},
  \bibinfo{author}{\bibfnamefont{H.}~\bibnamefont{Michinel}},
  \bibinfo{author}{\bibfnamefont{J.~I.}~\bibnamefont{Cirac}},
  \bibinfo{author}{\bibfnamefont{M.}~\bibnamefont{Lewenstein}},
  \bibnamefont{and} \bibinfo{author}{\bibfnamefont{P.}~\bibnamefont{Zoller}},
  \bibinfo{journal}{Phys. Rev. Lett.} \textbf{\bibinfo{volume}{77}},
  \bibinfo{pages}{5320} (\bibinfo{year}{1996}).
  
\bibitem{jaksch}
  D.~Jaksch, C.~Bruder, J.~I.~Cirac, C.~W.~Gardiner, and P.~Zoller, Phys. Rev. Lett. {\bf 81}, 3108 (1998).

\bibitem{pollack}
S.~E.~Pollack, D.~Dries, M.~Junker, Y.~P.~Chen, T.~A.~Corcovilos, and R.~G.~Hulet, Phys. Rev. Lett. {\bf 102}, 090402 (2009).


\end{thebibliography}
\end{document}